# SPS-Sintered NaTaO$_3$-Fe$_2$O$_3$ Composite exhibits Large Seebeck Coefficient and Electric Current

Wilfried Wunderlich [1)], Takao Mori[2)], Oksana Sologub[2)],

1) Tokai University, Fac.Eng., Material Science Dep., Kitakaname 4-1-1, 259-1292 Hiratsuka-shi, Japan
2) Nat. Inst. Mat. Sci. (NIMS), Int. Center Mat. Nanoarchitec. (MANA), Tsukuba, 305-0044, Japan

NaTaO$_3$– 50wt% Fe$_2$O$_3$ composite ceramics showed a large Seebeck voltage of -300 mV at a temperature gradient of 650 K yielding a constant Seebeck coefficient of more than -500μV/K over a wide temperature range. We report for the first time that SPS sintering at low temperature (870 K) could maintain the short-circuit current of -80 μA, which makes this thermoelectric material a possible candidate for high-temperature applications up to 1623 K. The reason for the good performance is the interface between Fe$_2$O$_3$ and surrounding NaTaO$_3$ perovskite. When spark-plasma sintering (SPS) is used, constitutional vacancies disappeared and the electric conductivity increases remarkably yielding *ZT* of 0.016.

Keywords: *Ceramic processing Seebeck coefficient, Energy harvesting*

**Introduction**

The pressing problem of CO$_2$ increase and climate change requires the search for new energy sources such as the thermoelectric power generators (TEG), which can turn waste heat into usable electricity when operating at high temperatures. The research for new thermoelectric ceramic materials began in the last decade and Nb-doped SrTiO$_3$ [1-2] NaCoO$_3$ [3], and CaCoO$_3$ [4] were found to have a remarkable figure of merit ZT. They are already successfully established in devices for high-temperature electric generators. A detailed band-structure study of the pervoskite-based Nb-doped SrTiO$_3$ material has emphasized the combination of large and small effective masses as the reason for the large Seebeck coefficient [5,6]. While Co-based perovskites [7-9] have been investigated also as potential thermoelectric materials, our search for new materials yielded to the NaTaO3 perovskite material, which is known as efficient photo catalyst for splitting water [10,11] and its large effective mass [5].

The composite material NaTaO$_3$-Fe$_2$O$_3$ shows a large Seebeck voltage of -300 mV at a temperature gradient of *ΔT* =650 K with linear temperature dependence [10-15] and is stable up to 1623 K [12, 15]. Yet its large resistivity has to be lowered for increasing the power factor and figure-of-merit. We have reported previous results on spark-plasma sintering experiments [15], where a remarkable increase in electric conductivity was achieved, but the Seebeck voltage has dropped. The conclusion was that either the interface structure or the microstructure has changed due to the high-temperature plasma, the vacuum, diffusion from the carbon crucible, or when quenching during fast cooling and are responsible for the decrease of the Seebeck coefficient. Further findings were that composites processed from Fe$_2$O$_3$ and NaTaO$_3$, or additions of NaFeO$_3$ deteriorate the electric conductivity and yield to an insulator. The reason for the good performance of this composite material is the interface between Fe$_2$O$_3$ and NaTaO$_3$ with perovskite structure. The largest Seebeck voltage was measured when the second phase Fe$_2$O$_3$ reaches an amount of 50 mol% [10], which is just the percolation limit when the second phase starts to surround the perovskite phase NaTaO3. Metallic behavior with high carrier concentration was recently found at similar NaTaO$_3$/SrTiO$_3$ perovskite interfaces [16].

Hence, the goal of this paper is to gain deeper insight in the materials behavior with the goal to improve both, Seebeck voltage and electric conductivity. For optimum densification a second sintering step is required after calcination and grinding. So, this paper describes processing of these composite ceramics on different routes and compare both, thermoelectric and microstructural properties of the resolved specimens.

**Experimental Procedure**
Powders in μm-size of NaTaO$_3$ and Fe (Fine Chemicals Ltd., Japan) were weighed according to the desired weight ratio of 50 mol-% Fe$_2$O$_3$ and mixed in a mortar for at least 10 min. Then the mixture was put in a steel cylinder with 15 mm diameter and cold-pressed with a stress of 50 MPa. These pellets were used in the following different synthesis methods as sketched in Fig. 1. The conventional calcination and sintering route is shown as paths (a) and (b) in Fig. 1 and details have been described in [12-14]. A sliced specimen with 10x2x2 mm dimensions was measured using the thermoelectric multi-measuring device ZEM3 (Ulvac Ltd., Japan) as described in the following section.

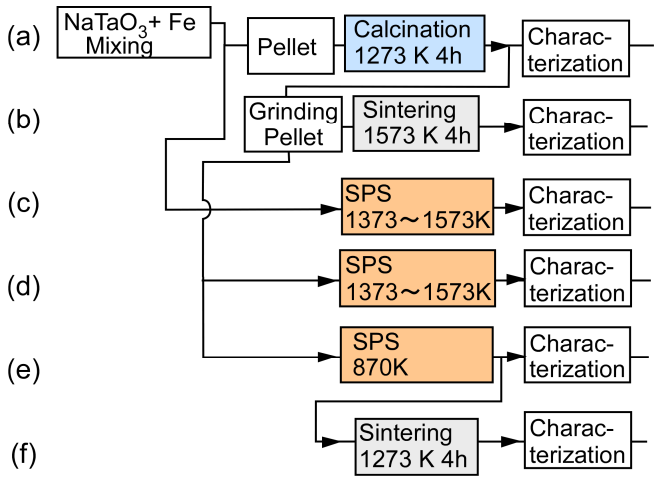

**Fig.1** The processing procedure yielded to different routes as indicated, (a) calcination, (b) conventional sintering, (c) SPS at 1373 to 1573 K, (d) calcination 1273 K in air, then SPS at 1373 K to 1573 K, (e) calcination 1273 K in air, SPS at 870 K, and (f) same as (e) with additional sintering at 1273 K for 4h.

The next straight-forward step is to try spark-plasma sintering (SPS) on the cold pressed pellet (route (c) in Fig. 1) [15]. The report on such specimens showed improved resistivity, but poor Seebeck voltage [15]. The present paper focusses on SPS sintering after calcination, then crushing, mixing the powder again and put it in a 15 mm graphite cylinder and attached in the Doctor Sinter 1080 SPS device (Syntex Sumitomo Heavy Industries, Ltd). Two regimes were tested and are marked as high and low temperature routes (d) and (e) in Fig. 1, namely 1373 to 1573 K and 870 K, respectively. The maximum pressure of 80 MPa was applied and kept constant, while temperature, spark plasma voltage and current were increased as described in detail in [15]. The plasma chamber was first evacuated, and then the sintering was performed at 1 atm Ar pressure. The duration of sintering was kept constant as 600 s.

The obtained specimens were characterized concerning their microstructure and composition by using a Hitachi 3200-N scanning electron microscope (SEM) operated at 20 kV and equipped with an electron dispersive spectrometer (EDS, Noran Ltd.). The thermoelectric voltage was measured against nickel wires with a distance of 10mm in a home-made device when applying a temperature difference between the micro-ceramic heater (Sakaguchi Ltd. MS 1000) up to 1273 K, and maintaining the cold end at around 473 K, as reported elsewhere [12-14]. The electric measurement devices (Sanwa PC510) recorded the data directly on a computer. The densities of the specimens were calculated from their mass-to-volume ratio, where the specimen dimensions were measured by a caliper.

## Results and Discussion

At first, a conventionally calcined and sintered specimen as described in [12] was measured for the first time by using a ZEM3. The thermoelectric measurements as displayed in Fig. 2 confirmed the relatively high resistivity, which decreases as a function of temperature.

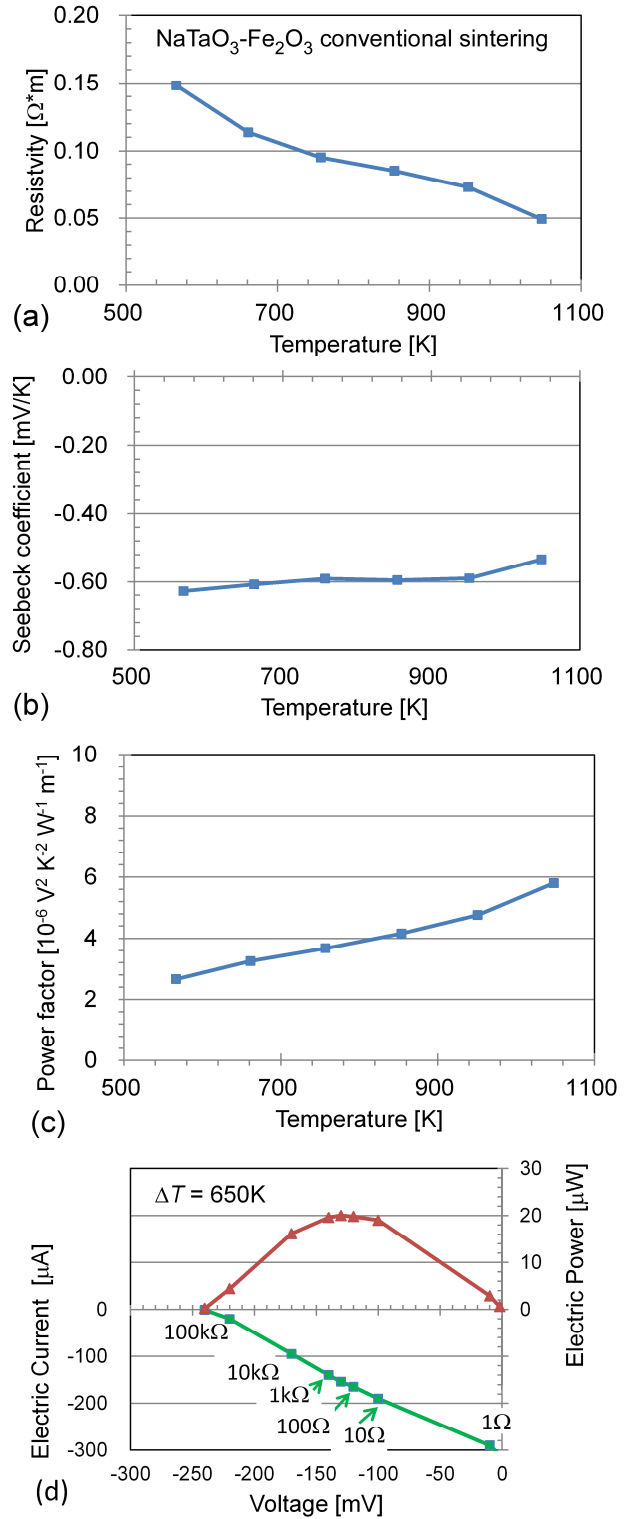

**Fig. 2** Thermoelectric properties of $NaTaO_3$-50mol%$Fe_2O_3$ produced by calcination and conventional sintering, (a) resistivity, (b) Seebeck coefficient, (c) power factor $S^2\sigma$, (d) closed circuit current as a function of the Seebeck voltage and generated electric power at $\Delta T$=650 K.

At $T$=1000 K the resistivity reached as less as 0.05 Ω m, as shown in Fig. 2 (a), about one order of magnitude better than previously reported values [12]. The Seebeck coefficient reaches -0.6 mV/K (Fig. 2 (b)) and is almost constant over a wide temperature range from 600 to 1000 K, allowing such specimen to be used as reference specimens for calibrating a measuring device. These measurements are in good agreement with previously published ones [12-14]. The power factor $S^2 \sigma$ as displayed in Fig. 2 (c) reaches a value of 6 $10^{-6}$ W/m K, which is yet one or two orders of magnitude smaller than comparable oxides [1-4]. The reason is that the resistivity remains still unacceptable large as concluded from this measurement, and motivates the SPS processing trial reported in the following section. When closing the open circuit with different load resistors (1 Ω to 100 kΩ), the electric current was measured. The maximum in this closed circuit current was -0.32 mA and the maximum Seebeck voltage was -240 mV with a linear dependence (Fig. 2 (d)). This relation yields at an intermediate operation point at $U$ = -130 mV, $I$ = -154 µA to a maximum power of $P = U*I$ = 20 µW for a temperature difference $\Delta T$ = 650K, which is still a factor of 10 below state-of-the-art oxides [3, 4]. By assuming a thermal conductivity of $\kappa$ = 1.7 W/m K such as typical for comparable perovskites, we end up with a figure–of-merit in the order of ZT = 0.016, still low, but promising.

In order to improve the electric conductivity, we tried the SPS sintering at high temperatures 1373, 1473 and 1573 K on previously calcined specimens as sketched in Fig. 1 (c). The absolute value of the Seebeck voltage was reduced to -38 mV as shown in Fig. 3 (a) and also a lower absolute value of the closed circuit current of -40 µA (Fig. 3 (b)). Compared to a maximum of the Seebeck coefficient of $S$ = -0.62 mV/K for the conventional sintered NaTaO$_3$-Fe$_2$O$_3$ samples, the Seebeck coefficient decreased to $S$ = -0.075 mV/K. The vanishing close circuit current can be explained by similar experience from a SPS sintering experiment without pre-calcining this material [15], route (c) in Fig. 1. One or more of the factors, vacuum, Ar pressure, spark plasma, the surrounding carbon container, or the fast cooling rate, are apparently bad conditions. which are deteriorating both, the Seebeck voltage and resistivity, probably by reducing the oxygen content or changing the interface structure.

Microstructural characterization by SEM, as shown in Fig. 4, confirmed that large pores due to insufficient compaction disappeared, but smaller pores on a sub-micrometer scale are still present in some areas. The shape and location between Fe$_2$O$_3$ (black) and NaTaO$_3$ (white) suggest that two mechanisms, the first solving Fe into NaTaO$_3$ perovskite lattice and the other oxidation of Fe to Fe$_2$O$_3$, both are responsible for these pores due to unbalanced stoichiometry. Nevertheless, the density has remarkably increased compared to the conventional sintering as shown in Table 1, which compares all experiments.

In order to improve the thermoelectric properties, we have processed another sample by calcination and subsequent SPS at low temperature of 870 K according to route (e) in Fig. 1, see also Table 1. This sample maintained the large Seebeck coefficient of -0.5 mV/K (Fig. 5 (a)). By subsequent sintering at 1273 K 4h (step(f)) simultaneously a maximum current of -80 µA was measured when closing the circuit with the load of a

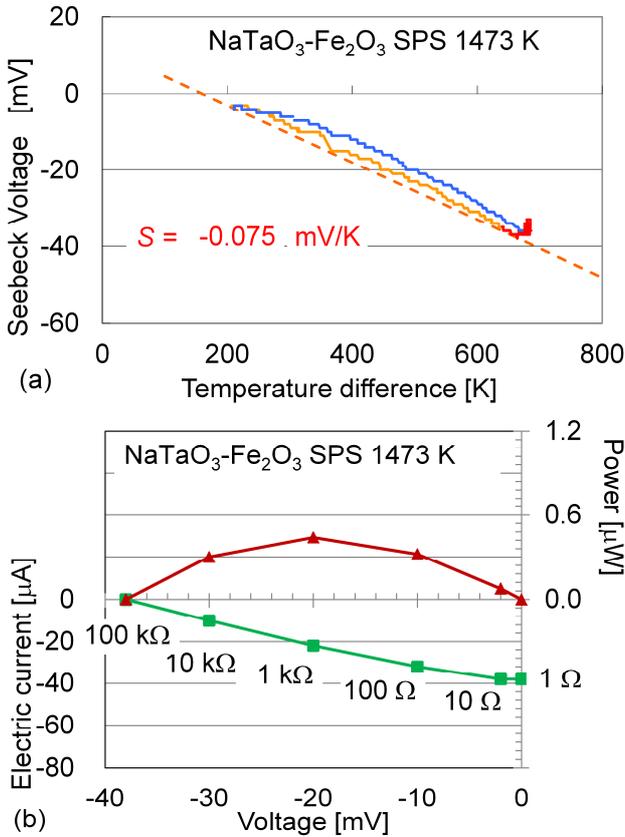

**Fig. 3** (a) Seebeck voltage of a NaTaO$_3$ -50mol% Fe$_2$O$_3$ specimen, which was calcined and then SPS sintered at 1473 K. (b) Electric current as a function of Seebeck voltage at ∆T = 650 K when load resistors are applied.

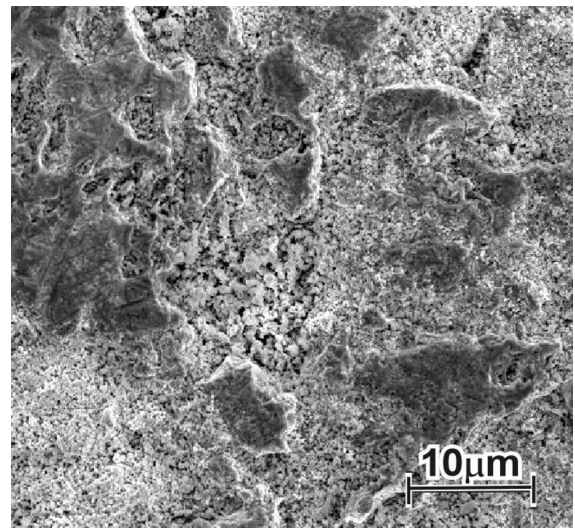

**Fig. 4.** Microstructure of the SPS specimen sintered at 1473 K

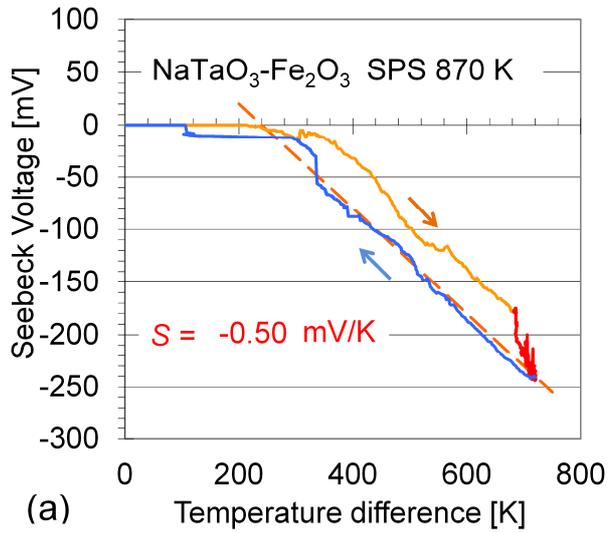

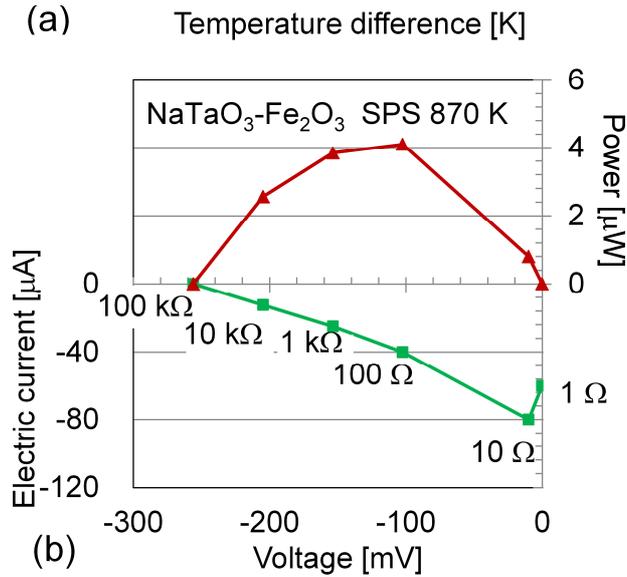

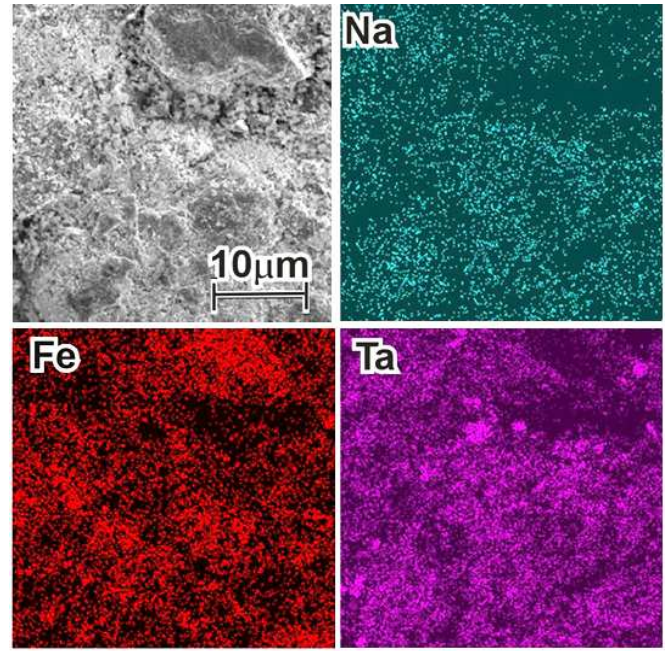

**Fig. 6** Element mapping by SEM-EDX of the specimen calcined and SPS sintered at 870 K.

**Fig. 5** Thermoelectric characterization of the NaTaO$_3$ -50mol% Fe$_2$O$_3$ calcined and SPS sintered at 870 K, and sintered at 1273 K 4h, (a) Seebeck voltage as function of a large temperature gradient, where the arrows mark the heating and cooling, and the slope marks the Seebeck coefficient. (b) Electric current and power as a function of Seebeck voltage at ΔT=650K, when load resistors are applied as marked.

10 Ω resistor. This is the first time that such specimens revealed good thermoelectric performance after SPS sintering, in spite of the fact that the output power is still one order of magnitude less than that of conventional sintered specimens, as shown in Fig. 5 (b). The specific resistivity decreased one order of magnitude and reached 0.015 Ω m at a temperature of 1100 K, which means the figure-of-merit is $ZT$ = 0.016. This fact confirms the benefits of using SPS sintering for closing pores and increasing density. In SEM micrographs, like that shown in Fig. 6, areas with stoichiometric frustration have almost vanished. The EDX mapping confirmed the homogenous distribution of chemical elements in the dark phase, Fe$_2$O$_3$, as well as in the bright one, NaTaO$_3$. The results of this study are summarized in table 1 which has the same notation of processing steps as Fig. 1. While solving of Fe into NaTaO$_3$ and calcination requires slow annealing (steps (a) and (b)), densification by maintaining a large negative Seebeck coefficient was confirmed here for the first time. The large electric conductivity is explained by the enlarged densification due to applied pressure during SPS sintering. On the other hand, the large thermo power can only be maintained, when the specimens are sintered by SPS at relatively low temperature below 900 K with subsequent sintering in air (steps (e) and (f)). This fact also confirmed the self-repairing ability of this composite besides the long-term stability due to the fact that thermoelectric properties improved after each sintering cycle [12, 13]. Further investigations should confirm the detailed mechanism why a large negative Seebeck coefficient and a fairly large electric conductivity are achieved at the same time. The fact that the optimum composition fits suppressed in the highly symmetric perovskite. In detail, well to the percolation composition [12], where the volume fraction of both phases are equal, as well as results from recent reports [2, 16], indicate that a confined two-dimensional electron gas (2DEG) at the heterogeneous interface is the reason for the good performance of this composite. Further improvement of ZT is expected from nano-scaled composites or superlattices or by finding suitable co-dopants.

**Table 1** Density and thermoelectric properties of the NaTaO$_3$ -50 mol% Fe$_2$O$_3$ composites as a function of processing method and sintering temperature

| Processing method | Sinter-Temperature [K] | Density [kg/m$^3$] | Seebeck coefficient [mV/K] | Electric current [μA] | Maximal Power [μW] |
|---|---|---|---|---|---|
| (a) [12] | 1273 | 2.55 | -0.50 | -50 | 2.5 |
| (b) [12,13], Fig 1 | 1473 | 2.82 | -0.62 | -320 | 20 |
| (c) [15] | 1473 | 4.36 | -0.001 | -10 | 0.01 |
| (d), Fig. 3 | 1473 | 4.43 | -0.075 | -40 | 0.5 |
| (e) | 870 | 2.88 | -0.50 | -10 | 0.1 |
| (f), Fig. 5 | 1273 | 3.56 | -0.50 | -80 | 4 |

**Conclusions**
The results of this spark-plasma sintering study on the $NaTaO_3$-$Fe_2O_3$ composite material confirmed the following facts.
(1) For the first time we could achieve a large negative Seebeck voltage of this n-type thermoelectrics and closed circuit current with remarkable power output by substantially decreasing the SPS sintering temperature from 1273 K to 870 K.
(2) The densities of the SPS-processed specimens were much higher than the cold pressed specimens with fewer pores and cracks, leading to a significantly lower electrical resistivity.
(3) Sub-micrometer sized pores due to constitutional frustration almost disappeared confirming the homogeneous microstructure after calcination and subsequent SPS sintering at low temperature. While densification improved the electric conductivity significantly, the large thermo power can only be maintained by slow heating in air, low SPS sintering temperature and subsequent sintering at high temperature.

**References**
[1] S. Ohta, T. Nomura, H. Ohta, H. Hosono, K. Koumoto, "Large thermoelectric performance of heavily Nb-doped $SrTiO_3$ epitaxial film at high temperature", Appl. Phys. Lett. 87 092108 (2005)
[2] H. Ohta, S. Kim, Y. Mune, T. Mizoguchi, K. Nomura, S. Ohta, T. Nomura, Y. Nakanishi, Y. Ikuhara, M. Hirano, H. Hosono, K. Koumoto, "Giant thermoelectric Seebeck coefficient of a two-dimensional electron gas in $SrTiO_3$", Nature Materials 6 129 (2007)
[3] I. Terasaki, Y. Sasago, and K. Uchinokura, "Large thermoelectric power in $NaCo_2O_4$ single crystals", Phys. Rev. B 56 R12685 (1997)
[4] R. Funahashi, I. Matsubara, H. Ikuta, T. Takeuchi, U. Mizutani, S. Sodeoka, "An Oxide single crystal with high Thermoelectric Performance in Air", Jpn. J. Appl. Phys 39 L1127 (2000).
[5] W. Wunderlich, H. Ohta, K. Koumoto, "Enhanced effective mass in doped SrTiO3 and related perovskites", Physica B 404 2202 (2009)
[6] K. Shirai and K. Yamanaka, "Mechanism behind the high thermoelectric power factor of $SrTiO_3$ by calculating the transport coefficients", J. Appl. Phys. 113 053705 (2013);
[7] P. Tomes, R. Robert, M. Trottmann, L. Bocher, M.H. Aguirre, A. Bitschi, J. Hejtmanek, A. Weidenkaff, "Synthesis and Characterization of New Ceramic Thermoelectrics Implemented in a Thermoelectric Oxide Module", J. Elec. Mat. 39 [9] 1696 (2010)
[8] W. Wunderlich, H. Fujiwara, "The Difference between thermo- and pyroelectric Co- based RE-( = Nd, Y, Gd, Ce)-oxide composites measured by high-temperature gradient", J. Electronic Materials 40 [2] 127-133 (2011)
[9] W. Wunderlich, Large Seebeck Voltage of Co, Mn, Ni, Fe- Ceramics, Advances in Ceramic Science and Engineering (ACSE) 2 [1] (2013) 9-15, http:// www.acse-journal.org/Download.aspx?ID=5579
[10] H. Kato, A. Kudo, "New tantalate photocatalysts for water decomposition into H and $O_2$", Chemical Physics Letters 295 487–492 (1998)
[11] A. Kudo, and H. Kato, "Effect of lanthanide-doping into $NaTaO_3$ photocatalysts for efficient water splitting", Chemical Physics Letters 331 373-377 (2000)
[12] W. Wunderlich, "$NaTaO_3$ composite ceramics - a new thermoelectric material for energy generation", J Nucl. Mat. 389 [1] 57-61 (2009)
[13] W. Wunderlich, S. Soga, "Microstructure and Seebeck voltage of $NaTaO_3$ composite ceramics with additions of Mn, Cr, Fe or Ti", J. Cer. Proc. Res. 11 233-236 (2010)
[14] W. Wunderlich, B. Baufeld, "Development of Thermoelectric materials based on $NaTaO_3$ –composite ceramics", 1-27, in "Ceramic Materials", ed. W. Wunderlich, Intech Publisher, doi: 10.5772/243 1-27 (2010)
[15] W. Wunderlich, T. Mori, O. Sologub, B. Baufeld, "SPS-Sintering of NaTaO3-Fe2O3 Composites", J. Australian Ceramic Society 47 [2] 57-60 (2011).
[16] S. Nazir and U. Schwingenschlögl, "High charge carrier density at the $NaTaO_3$/$SrTiO_3$ hetero-interface", Appl. Phys. Lett. 99 073102 (2011), doi:10.1063/1.3625951.